# Digital instability of a confined elastic meniscus

John S. Biggins[a,b], Baudouin Saintyves[c], Zhiyan Wei[a], Elisabeth Bouchaud[c,d], and L. Mahadevan[a,e,f,g,1]

[f]Wyss Institute for Biologically Inspired Engineering, [g]Kavli Institute for Bionano Science and Technology, [a]School of Engineering and Applied Sciences, and [e]Department of Physics, Harvard University, Cambridge, MA 02138; [b]Cavendish Laboratory, University of Cambridge, Cambridge CB3 0HE, United Kingdom; [c]Commissariat à l'Energie Atomique, Institut Rayonnement Matière de Saclay, Service de Physique de l'Etat Condensé, F-91191 Gif-sur-Yvette Cedex, France; and [d]Ecole Supérieure de Physique et de Chimie Industrielles ParisTech, Unité Mixte de Recherche Gulliver, 75231 Paris Cedex 5, France

Thin soft elastic layers serving as joints between relatively rigid bodies may function as sealants, thermal, electrical, or mechanical insulators, bearings, or adhesives. When such a joint is stressed, even though perfect adhesion is maintained, the exposed free meniscus in the thin elastic layer becomes unstable, leading to the formation of spatially periodic digits of air that invade the elastic layer, reminiscent of viscous fingering in a thin fluid layer. However, the elastic instability is reversible and rate-independent, disappearing when the joint is unstressed. We use theory, experiments, and numerical simulations to show that the transition to the digital state is sudden (first-order), the wavelength and amplitude of the fingers are proportional to the thickness of the elastic layer, and the required separation to trigger the instability is inversely proportional to the in-plane dimension of the layer. Our study reveals the energetic origin of this instability and has implications for the strength of polymeric adhesives; it also suggests a method for patterning thin films reversibly with any arrangement of localized fingers in a digital elastic memory, which we confirm experimentally.

hysteresis | confinement | elastomer | gel

In adhesive joints, the strains and stresses due to joint loading are magnified by the effects of geometric confinement and scale separation (1), making them susceptible to stress-driven instabilities that often lead to failure. Joints usually fail in one of two broad ways: via adhesive failure along the solid–solid interface (2–5) or via bulk cohesive failure of the glue joint via cavitation (6–8). Though these modes of failure have been well documented and studied (see ref. 1 for a review), there is a third mode of failure, where an elastic instability at the meniscus may lead to fracture in its vicinity and can arise either when a joint is loaded under tension (9) or by a fluid that is injected into a cavity in the confined elastic layer (10). This mode of failure has been largely overlooked experimentally and is not understood theoretically. Interestingly, the last experiment is an elastic analog of a well-studied classical hydrodynamic free-surface instability associated with the relative motion between liquids of different viscosities in a narrow gap (11, 12), and provides a point for comparison. As we will see, the elastic instability is fundamentally different given its reversible nature and lack of dependence on interfacial forces. We use a combination of theory, experiment, and computation to unravel the mechanism behind the elastic meniscus instability, the threshold strain for its onset, the critical wavelength of the resulting fingers, and the nonlinear development of its amplitude.

Geometrically, our setup, sketched in Fig. 1A, consists of a thin, highly elastic layer occupying the region $-a/2 < z < a/2$, $-\infty < x < \infty$, $0 < y < l$ with $a/l \ll 1$ that is adhered to rigid plates at $z = \pm a/2$. Experimentally, we used a layer of polyacrylamide gel with a shear modulus of 550 Pa, thickness $a \in [0.28, 10.64]$ mm, and width $l \in [50, 60]$ mm bound between 10-mm-thick glass plates that were ~200 mm long. The plates are then pulled apart, increasing their separation to $a + \Delta z$, while maintaining adhesion. Experimentally, the separation was increased at a constant speed of ~200 mm/s. As the rigid plates are separated, the free boundaries of the elastomer (at $y = 0$ and $y = l$) retreat to form an elastic meniscus that is curved in the direction perpendicular to the plates but remains parallel to its original position, thus penetrating into the elastic film without causing any loss of adhesion to the glass plates. At a critical separation of the plates, this curved meniscus loses stability via a sharp transition to an undulatory configuration in which fingers of air protrude into the elastomer, shown schematically in Fig. 1A. To ensure that elastic equilibrium was achieved at each stage, and to rule out any rate dependence, we also performed experiments at much lower velocities and saw quantitatively similar results.

Fig. 1B shows the undulatory pattern observed. We note that this instability is qualitatively different from the crack-like adhesive undulatory instabilities seen at the glass–gel contact line when adhesion starts to fail (2). In our experiments, adhesion is maintained everywhere due to the natural propensity of polyacrylamide to stick strongly to glass. Thus, fingers appear along the retreating elastic meniscus. Fig. 1B also shows a loading/unloading hysteresis loop for the transition, showing that the instability sets in suddenly past a given threshold in displacement via a subcritical instability, leading to large amplitude "digits" or fingers whose amplitude grows further upon further loading (Movie S1). On unloading, the fingers snap back at a lower value of the displacement, suggestive of the hysteretic nature of this first-order transition (Movie S2). We find that the undulatory transition is fully reversible and has no dependence on the shear modulus of the elastomer, strongly suggesting that the phenomenon is purely elastic. The similarity in the small smooth part of the loading and unloading curves, which corresponds to a plate separation of ~1.5% at most, is likely due to inhomogeneities in the meniscus when it was first formed via our molding protocol. To test this, we waited for up to 30 min after the destabilization of the front, and did not see any additional fingers form. On retracing the loading/unloading cycle, we saw that the system traced the same curves as the first time, consistent with this explanation. Finally, we performed identical experiments in oil rather than in air to determine the effect of surface tension on the instability (Movie S3), and find that the system responds just as when it is in air, eliminating a role for the effects of surface tension in the phenomenon. It is useful to contrast these observations with the case of viscous fingering (11), where fingering is dynamic and out of equilibrium, and surface tension effects cannot be neglected.

Because the deformations involved are large, we resorted to numerical simulations of the process in terms of a finite element method, using an incompressible neo-Hookean constitutive model for the elastic layer. To capture the subcritical nature of the instability, we needed to carry out a dynamical simulation

APPLIED PHYSICAL SCIENCES

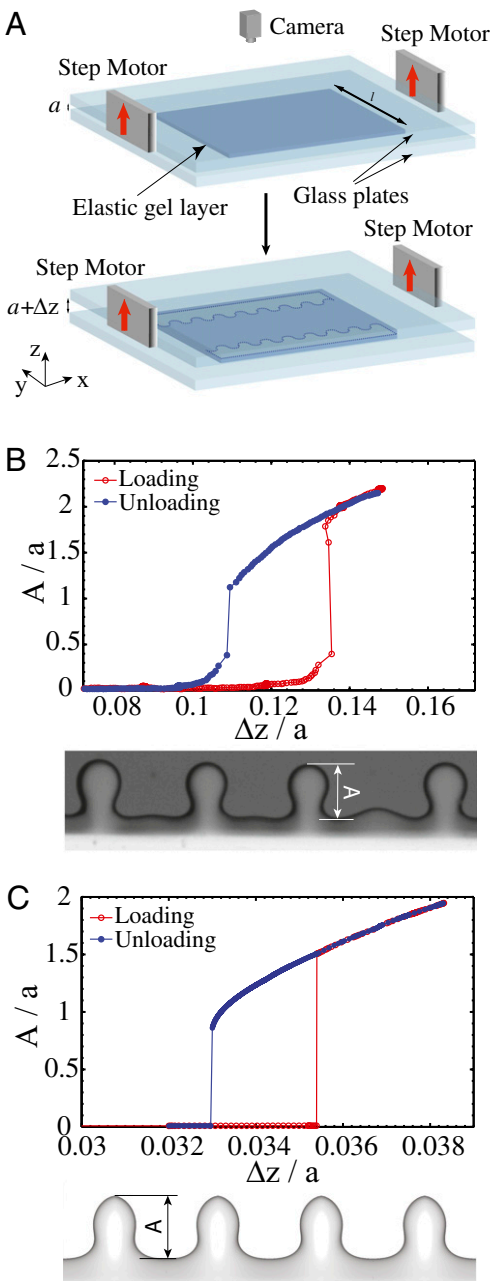

**Fig. 1.** (A) Schematic of an elastic layer between two rigid plates. Pulling the plates apart causes the two free menisci to lose stability by forming a series of undulating digital fingers. (B) The experimentally measured amplitude of the fingers as a function of plate separation $\Delta z$, along with a top view of the undulating meniscus showing the fingers of air (*Lower*) invading the elastic layer (*Upper*). The layer thickness is $a = 3.05$ mm and the width is $l = 56$ mm. Observe the hysteresis in the transition associated with the difference between the loading and unloading curves. (C) Numerical results for an identical quasi-static loading and unloading protocol (*SI Appendix, Numerical Simulations*) calculated using a finite element method for an elastic layer of thickness $a = 1$ mm and width $l = 40$ mm show the same qualitative features—a hysteretic transition (*Upper*) associated with the formation of undulating fingers (*Lower*).

with damping, modeled via a Rayleighian dissipation function. The boundary conditions were imposed using the symmetry of the problem, and a small amount of initial noise in the position of the meniscus was used to seed the instability. All numerical simulations were carried out using a commercial finite element package ABAQUS (*SI Appendix, Numerical Simulations*), allowing us to reproduce this instability. We note that once the fingers develop and have a finite amplitude, they may not be described by a single-valued function, but this is not an issue in our simulations, which use a natural coordinate system for the meniscus. In Fig. 1C, we show that both the form of the fingers and the hysteresis loop associated with a loading/unloading loop arise in a purely elastic simulation.

To understand our experiments and numerical simulations, we start by estimating the energies and length-scales in the problem. Displacement of a point in the central plane of the elastomer by an amount $u \sim a$ in the $y$ direction of the $x-y$ plane leads to a shear strain in the elastomer $\gamma \sim u/a$. Because the subcritical fingering transition is purely elastic, it is likely to occur at large strains with a threshold $\gamma \sim 1$ when geometrically nonlinear effects are important. Incompressibility of the elastomer implies that $\Delta z l \sim u a$, so that the instability threshold $\Delta z_t \sim O(a^2/l)$, which vanishes for infinitesimally thin films when $l/a \to \infty$. We note that this threshold arises from purely geometric considerations and expect that it does not depend on any material properties, because the only energy scale in the system, the shear modulus, can be scaled away. Furthermore, if fingers form with wavelength $\lambda$ and amplitude $A$, this introduces additional strain associated with the in-plane distortion of magnitude $A/\lambda$. At the onset of the instability, the elastic screening length $O(a)$ must scale with the thickness of the layer, so that we expect the wavelength of the instability to also be independent of any material parameters, with $\lambda \sim O(a)$. However, how is it that the formation of fingers, which are areas that have receded deeply into the bulk and therefore undergone huge shear strains, can reduce the total shear energy in the elastomer?

To clarify how fingering can alleviate shear, we first build a very simple model completely neglecting in-plane strain. Again, focusing our attention on the central plane of the elastomer, we assume that it is made of thin strips of width $dx$, which we treat as elastically independent. If one of these strips is stretched in-plane by a factor $\lambda_y$ in the $y$ direction and a factor $\lambda_x$ in the $x$ direction, as shown in Fig. 2 A and B, the small thickness of the strip guarantees that the displacements in the $x$ direction are small compared with $a$, and therefore do not give rise to large shear strains. However, a point with coordinate $y$ is moved by an amount $(y-l/2)(1-\lambda_y)$ and so suffers a strain $\gamma \sim (y-l/2)(1-\lambda_y)/a$, and the elastic shear energy of the strip is therefore $E_s = \propto \int_0^l \gamma^2 dy \propto (1-\lambda_y)^2$. A stretch in the $z$ direction by a factor of $(1 + \Delta z/a)$, together with volume conservation requires $(1+\Delta z/a)\lambda_x \lambda_y = 1$, which allows us to rewrite the shear energy of our strip as $E_s \propto (1+\Delta z/a - 1/(\lambda_x))^2$. Plotting this as a function of $\lambda_x$ in Fig. 2C, we see that the energy has a minimum at $\lambda_x = 1/(1+\Delta z/a) < 1$ for $\Delta z > 0$. However, because our system is infinite in the $x$ direction, we know that the average $x$-stretch $\langle \lambda_x \rangle = 1$; otherwise, the strips will build up infinite displacements in the $x$ direction. Inspecting Fig. 2C, we see a large nonconvex region extending from the minimum till $\lambda_x \to \infty$, i.e., the total energy of the system is minimized when $\Delta z > 0$ with most strips being stretched by the optimal value of $\lambda_x = 1/(1+\Delta z/a) < 1$ and a very small number having large $\lambda_x$; these digits dig deep into the bulk of the elastomer, leading to the fingering instability. We note that if the energy was convex, the minimum energy compatible with the average stretch $\langle \lambda_x \rangle = 1$ would be achieved by each strip individually taking $\lambda_x = 1$. This simple explanation thus accounts for how the lack of convexity drives the energetics of finger formation and predicts a first-order transition to a large-amplitude state, consistent with the experimentally observed hysteresis shown in Fig. 2B.

Though our zero-dimensional model provides a mechanism for the instability, it is unable to provide information about the wavelength and threshold for the instability; for this, we now turn to an asymptotic simplification of the 3D problem by taking advantage of the small thickness and symmetry of the elastic layer. We expand the displacement vector $\mathbf{U}(x,y,z)$ to leading order in $z$ and impose the condition that $\mathbf{U} = \pm \Delta z \hat{\mathbf{z}}/2$ at $z = \pm a/2$, leading to the form

$$\mathbf{U} = (1-2z/a)(1+2z/a)\mathbf{u}(x,y) + (z\Delta z/a)\hat{\mathbf{z}}, \quad [1]$$

where $\mathbf{u}(x,y)$ is the 2D displacement of a point on the central ($z=0$) plane. With $\nabla$ as the in-plane gradient operator and $I$ as the 2D identity matrix, we can then write the 3D deformation gradient, $F_{ij} = \delta_{ij} + \partial_j U_i$, as

$$F = I + (1-4z^2/a^2)\nabla\mathbf{u} - 8z\mathbf{u}\hat{z}/a + (1+\Delta z/a)\hat{z}\hat{z}, \quad [2]$$

and see the decomposition that results as a consequence of scale separation.

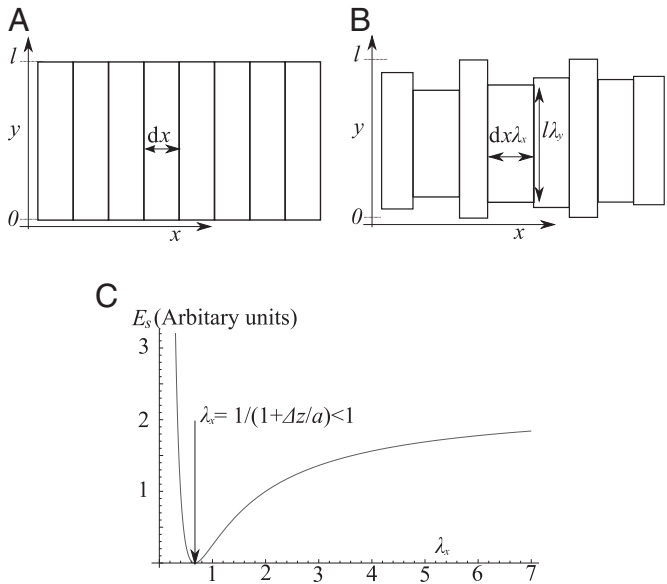

**Fig. 2.** (*A*) The simplest model that characterizes the phenomenon focuses on the central ($z=0$) plane of the elastomer and considers it to be composed of many independent thin strips of width d$x$. (*B*) A schematic of the deformation of the strips when stretched perpendicular to the plane of the paper leads them to undergo independent planar deformations. (*C*) The constraint of incompressibility causes the energy of a strip $E_s$ to not be a convex function of $\lambda_x$ so the minimal average energy with $\langle\lambda_x\rangle = 1$ is achieved by most strips taking the optimal value of contraction and a small number taking divergent values and hence receding deeply into the bulk and forming fingers. This minimal model highlights the mechanism of instability but provides no information about the wavelength and threshold for the instability (see text).

To characterize the energetic cost of this deformation, we model the elastomer as an incompressible neo-Hookean solid with volumetric elastic energy density $\frac{1}{2}\mu\mathrm{Tr}(F.F^T)$, which we can explicitly integrate in the thickness direction. Here we assume that surface tension effects are unimportant, as our experiments show. Thus, when the energy of the system is rescaled by this single constant, what remains is a purely geometric problem. The constraint of volume preservation in the elastomer when integrated through the depth requires us to introduce a 2D pressure field $P(x,y)$ that constrains the depth-averaged volume change at each point in the elastomer, and leads us to an effective 2D energy density $L$:

$$L(\mathbf{u},P) = \mu \int_{-a/2}^{a/2} \frac{1}{2}\mathrm{Tr}(F.F^T) - \frac{P(\mathrm{Det}(F)-1)}{1+\Delta z/a} dz$$

$$\propto \frac{1}{2}\mathrm{Tr}(G.G^T) + \frac{16}{5}|\mathbf{u}/a|^2 - P\left(\mathrm{Det}(G) - 1 + \frac{6}{5}\Delta z/a\right). \quad [3]$$

In carrying out the integral (*SI Appendix, Theoretical Model*), we have introduced an effective 2D deformation gradient $G = I + \frac{4}{5}\nabla\mathbf{u}$ and, because we expect $\Delta z \sim a^2/l \ll a$, retained only the leading-order term in $\Delta z/a$. We note that $\frac{\partial L}{\partial \nabla \mathbf{u}} = \frac{4}{5}(G - P\mathrm{Det}(G)G^{-T})$, so extremizing this energy leads to the following Euler–Lagrange equations for the planar displacement field $\mathbf{u}$ and the pressure $P$,

$$\frac{4}{5}\nabla^2\mathbf{u} - \mathrm{Det}(G)G^{-T}\cdot\nabla P = 8\mathbf{u}/a^2$$

$$\mathrm{Det}(G) = 1 - \frac{6}{5}\Delta z/a. \quad [4]$$

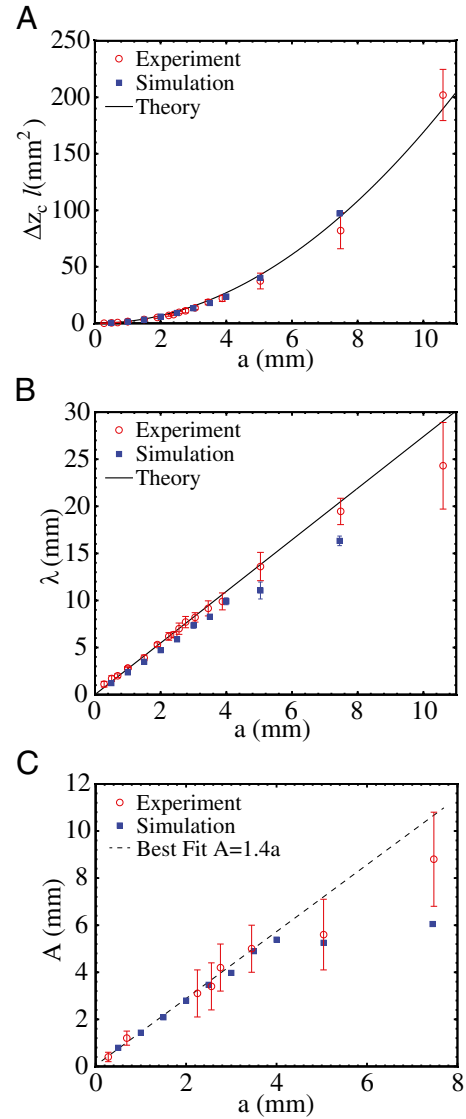

**Fig. 3.** Comparison of experimental and theoretical/numerical predictions. (*A*) Threshold separation $\Delta z_c \times$ width $l$ as a function of thickness $a$ shows that the experimental and numerical results follow the theoretical prediction (Eqs. **13** and **14**). (*B*) Finger wavelength $\lambda$ at instability as a function of thickness $a$ shows that the experimental and numerical results follow the theoretical prediction (Eqs. **13** and **14**). (*C*) Finger amplitude $A$ just after threshold as a function of thickness $a$ shows that the experimental and numerical results agree, but only over a range of thickness values. For large $a$, the separation of scales between the thickness $a$ and width of the film $l$ is less, and the number of wavelengths in the sample is smaller, leading to end effects that make agreement between theory and experiment only qualitative.

It is interesting to note that the form of the depth-integrated Eq. **4** is similar to the Darcy–Brinkman equation for flow through a dilute porous media (13), with the displacement reinterpreted as a velocity. Here, the most interesting aspect of the equation is the appearance of the bare displacement of the central plane **u**. On the free surfaces $y = 0, l$, we must satisfy the natural boundary condition

$$(G - P \operatorname{Det}(G) G^{-T}) \cdot \hat{\mathbf{y}} = 0. \quad [5]$$

Taking each field to be the sum of a large translationally invariant base state corresponding to the deformations before the instability and an infinitesimal oscillation in the $x$ direction, we may write

$$\mathbf{u} = Y_1(y)\hat{\mathbf{y}} + \epsilon \cos(kx) Y_2(y)\hat{\mathbf{y}} + \epsilon \sin(kx) X_2(y)\hat{\mathbf{x}} \quad [6]$$

$$P = 1 + P_1(y) + \epsilon \cos(kx) P_2(y). \quad [7]$$

Substituting this into [4]– [5] and solving for the translationally invariant fields, we get

$$Y_1(y) = \frac{3}{4} \Delta z (l - 2y)/a \quad [8]$$

$$P_1(y) = 6y\Delta z(y - l)/a^3 - \frac{6}{5}\Delta z/a. \quad [9]$$

At order $\epsilon$, the Euler–Lagrange equations (Eq. **4**) can be solved algebraically for $X_2$ and $P_2$ to yield a linear fourth-order eigenvalue equation for $Y_2$, whose solution provides us the wavelength and threshold for instability (*SI Appendix, Theoretical Model*). A substantial simplification arises by considering the limit $l \gg a$ and, consequently, $\Delta z \ll a$, which allows us to drop all terms proportional to $\Delta z$ except those also containing powers of $l$, reducing the final equation to

$$(a^2 k^2 + 10) a^2 k^2 Y_2(y) + a^4 Y_2^{(4)}(y) = 2(a^2 k^2 + 5) a^2 Y_2''(y), \quad [10]$$

which has the allowable decaying solutions

$$Y_2 = c_1 \exp\left(-\sqrt{10/a^2 + k^2}\, y\right) + c_2 \exp(-ky). \quad [11]$$

Substituting this into Eq. **5** gives $c_1 = -c_2 k^2/(5/a^2 + k^2)$ and a condition which yields the threshold separation $\Delta z_t$ for instability at wavenumber $k$

$$\Delta z_t = \frac{a^2}{l} \frac{\left(a^2 k^2 - \sqrt{a^2 k^2 + 10}\, ak + 10\right) a^2 k^2 + 25}{15ak}. \quad [12]$$

Minimizing this expression with respect to $k$ yields the threshold of the first unstable mode $(\Delta z_t)$ and the wavelength $(\lambda)$, which are given by

$$\Delta z_t \approx 1.69 a^2/l \quad [13]$$

$$\lambda \approx 2.74 a. \quad [14]$$

We see that the wavelength of the instability scales with the thickness of the elastic layer and the threshold displacement is inversely proportional to the slab width, and are independent of any material parameters, as we argued earlier based on scaling arguments.

In Fig. 3, we show a comparison of these predictions with experiments and numerical simulations (*SI Appendix, Numerical Simulations*); the results compare very well. Although our linearized analysis cannot extend beyond the point of instability, our finite element simulations have no such limit. Experimental and numerical results show that the amplitude of the fingers $A \approx 1.4a$, and confirm the subcritical nature of the instability with a region of bistability wherein the homogeneous and undulatory phases of the interface coexist. In this regard, our elastic instability is fundamentally different from the hydrodynamic Saffman–Taylor instability that is supercritical.

Our study has uncovered the form and nature of the confined elastic meniscus fingering instability in a minimal rectilinear setting using a combination of theory, experiment, and numerical simulation. We show the origin of the transition is essentially geometric and hence likely to be ubiquitous, just as its fluid counterpart is, and predict and verify the wavelength and threshold of the instability. At a practical level, our results have implications for the strength of elastic adhesive layers; because the peak strain jumps very significantly during the fingering transition, fingering is very likely to lead to fracture and adhesive failure. From our 2D model, the stored energy per unit area scales as $\mu a (\Delta z l/a^2)^2$, so that the normal stress that must be applied to the plates $\sigma_t \sim \mu l/a$, and predicts that the fracture stress of polymeric adhesives is inversely proportional to the thickness of the layer, and that the total strength of the adhesive bond increases faster than the adhesion area.

We have also shown that the transition is sudden with a region of bistability between the fingered and flat states. The hysteretic nature of the transition permits control over the placement of fingers or digits; if the system is in the bistable regime one may "write" out arbitrary "bits" onto the interface by applying a large perturbation at the desired location (Movie S4). These bits are completely reversible localized elastic structures, so that this fingering transition might be used to build a digital mechanical memory. Because our system produces fingers with wavelength proportional to the smallest length-scale in the problem—namely, the thickness of the layer—without any prepatterning on this length-scale, this mechanism may also have uses in microfabrication. Although the digitization instability is fully reversible, it may be easily made permanent by further cross-linking; additionally, the use of a nematic elastomer may allow the transition to be driven by heat or light rather than separation.

**ACKNOWLEDGMENTS.** This work was supported by the Royal Society (J.S.B.), the French ANR F2F project (E.B.), the MacArthur Foundation (L.M.), and Grant DMR0820484 from Harvard-MRSEC NSF (Materials Research Science and Engineering Centers, National Science Foundation).